# THREE-BODY DECAYS OF THE SUSY HIGGS BOSONS*

**Jan Kalinowski**

*Institute of Theoretical Physics, Warsaw University, ul. Hoża 69*
*00681 Warsaw, Poland*
E-mail: kalino@fuw.edu.pl

and

*Dipartimento di Fisica Teorica, Università di Torino, via P. Giuria 1*
*10125 Torino, Italy*

ABSTRACT

In this talk I discuss three-body decays of the SUSY Higgs bosons which proceed via intermediate two-body states in which one of the particles is off-mass-shell. Some of these decays will be shown to have substantial branching ratios in some range of the SUSY parameter space.

## 1. Introduction

The Higgs sector of the standard model (SM) still remains to be tested experimentally. Much effort has been devoted to Higgs boson searches in the past years and it remains to be one of the major physics goals of present and future experiments at high-energy colliders.[1] Not only we will have to find the Higgs boson(s) and measure its mass but it will be extremely important to measure its couplings to other particles. This will provide a testing ground for the idea of spontaneous symmetry breaking mechanism and to reveal the nature of the underlying theory as being the minimal standard model or one of its possible extensions, like for example supersymmetry.

From the extensive studies of the Higgs sector we have learned that almost all Higgs boson production channels and decay modes are plagued by large SM backgrounds. Therefore it is important to have precise calculations of the production cross sections and decay rates. Since the Higgs boson couplings to SM particles are proportional to their masses (modulo enhancement/suppression factors in extended

*Presented at the International Symposium and Workshops on Particle Theory and Phenomenology, May 17–26, 1995, Ames (Iowa).

models), the important decay modes are two-body decays to heaviest pair of particles allowed by the phase space. However, in the SM we have also learned that there are important cases when the so called below-threshold, or 3-body (or even 4-body), decay modes are in some cases also very important. A well known example[2] is the decay to gauge bosons $H_{SM} \to V^{(*)}V^{(*)}$, $(V = W, Z)$ below the threshold ($M_{H_{SM}} < 2M_V$) with one (or two) $V$'s being off-shell and decaying to fermions. Although suppressed by the off-shell propagator(s), this process is enhanced by the strong Higgs coupling to gauge bosons leading to appreciable branching ratios. In fact, the below-threshold 3-body decay $M_{H_{SM}} \to ZZ^* \to Zll$ $(\to 4l)$ provides a very important signature for the intermediate-mass Higgs boson with mass in the range $\sim 130GeV < M_H < 2M_Z$.

In the minimal supersymmetric extension of the standard model (MSSM) the decay pattern of Higgs bosons is much richer and 3-body decays may be even more important phenomenologically. In this talk I will discuss various three-body decay modes of the neutral and charged Higgs bosons in the MSSM. To obtain the branching ratios for these processes exact matrix-element calculations for the complete process have been performed, for example $H \to W^+ \; f\bar{f}'$ below $WW$ and $H \to t \; W^-\bar{b}$ below $t\bar{t}$ thresholds, or $H^+ \to \bar{b} \, W^+ b$ below $t\bar{b}$ threshold. The results presented here have been obtained in collaboration with A. Djouadi and P. Zerwas and details of the calculations can be found in Ref.[3] Similar study has been performed in a recent paper by Moretti and Stirling.[4]

## 2. Three-body decays in the MSSM

In minimal supersymmetric standard model the Higgs sector has a rich structure.[5] There are 5 physical Higgs bosons: two CP-even neutral ($h$, $H$, with $M_h < M_H$), one CP-odd neutral ($A$) and a pair of charged ($H^\pm$). Besides their masses, two additional parameters define the properties of these particles: a mixing angle $\alpha$ in the neutral CP-even sector and the ratio of the two vacuum expectation values $\tan\beta$, which from GUT restrictions is assumed to be in the range $1 < \tan\beta < M_t/M_b$, with the lower or the upper range, i.e. $\tan\beta \sim 1$ or $\tan\beta \sim M_t/M_b$, being favored since they provide an explanation of the large value of the top quark mass.

Supersymmetry leads to several relations among the parameters of the Higgs sector and only two of them are in fact independent, for example $M_A$ and $\tan\beta$. The relations among parameters are complicated by radiative corrections[6] as the top quark mass is large.[7] Since we are not interested here in detailed studies of effects of all possible higher-order corrections but only to find out which three-body decay modes are important, we will incorporate these radiative corrections in terms of the parameter

$$\epsilon = \frac{3\alpha}{2\pi s_W^2 c_W^2 \sin^2\beta} \frac{M_t^4}{M_Z^2} \log\left(1 + \frac{M_S^2}{M_t^2}\right), \tag{1}$$

$s_W^2 = 1 - c_W^2 \equiv \sin^2\theta_W$, neglecting non-leading effects due to non-zero values of the supersymmetric Higgs mass parameter $\mu$ and of the trilinear couplings $A_t$ and $A_b$ and assuming a universal soft supersymmetry-breaking scale $M_S$. Then the masses of the other Higgs bosons and the mixing angle $\alpha$ are given by

$$
\begin{aligned}
M^2_{h,H} &= \frac{1}{2}\Big[M_A^2 + M_Z^2 + \epsilon - \{(M_A^2 + M_Z^2 + \epsilon)^2 \\
&\quad -4M_A^2 M_Z^2 \cos^2 2\beta - 4\epsilon(M_A^2 \sin^2\beta + M_Z^2\cos^2\beta)\}^{1/2}\Big] \qquad (2)\\
M^2_{H^\pm} &= M_A^2 + M_W^2 \qquad (3)\\
\tan 2\alpha &= \tan 2\beta \frac{M_A^2 + M_Z^2}{M_A^2 - M_Z^2 + \epsilon/\cos 2\beta} \;, \quad [-\frac{\pi}{2} < \alpha < 0] \qquad (4)
\end{aligned}
$$

The Higgs boson couplings to fermions and gauge bosons are functions of the mixing angles.[5,8] These couplings, and hence the value of $\tan\beta$, determine to a large extent the decay pattern of the supersymmetric Higgs bosons. They turn out to be quite different for large or small $\tan\beta$.

For large values of $\tan\beta(\geq 5)$ the pattern is relatively simple because of strong enhancement of the Higgs couplings to down-type fermions. The neutral Higgs bosons will decay into $b\bar{b}$ and $\tau^+\tau^-$ pairs, the branching ratios being approximately equal to $\sim 90\%$ and $\sim 10\%$ respectively. The charged Higgs particles decay into $\tau\nu_\tau$ pairs below and into $tb$ above the top-bottom threshold. Only in two cases the three-body decays turn out to be interesting: (a) in the case of $h$ when its mass is close to its maximal value, and (b) in the case of $H$ when its mass is close to its minimal value; both exceptions occur however in a very limited domain of the parameter space. The respective Higgs bosons ($h$ in (a), and $H$ in (b)) have then (almost) SM couplings[a] and like in the SM the three-body decays into $WW^*/ZZ^*$, with one of $W/Z$ being off-shell occur with an appreciable rate. Since these processes have been extensively studied in the literature we will not discuss the large $\tan\beta$ case here (for a full discussion we refer to Ref.[3]).

This simple pattern can become complicated if one allows for the possibility of Higgs boson decays into supersymmetric particles, like charginos/neutralinos and sfermions.[8] Since three-body decays are of importance only for relatively light Higgs bosons, these decays will not be considered here.

For small values of $\tan\beta \sim 1$ the decay pattern is much richer due to several factors: (a) the couplings to bottom quarks and $\tau$ leptons are not significantly enhanced compared to their SM values, (b) the couplings to gauge bosons are not very suppressed, and (c) for masses below $\sim 2M_t$ for the heavy neutral and $\sim M_t$ for the charged Higgs bosons, the overwhelming decays into top quark channels are not yet available. Therefore a number of interesting three-body decay modes becomes available which we will discuss in some detail below. In Fig.1-3 we present the results for a representative value of $\tan\beta = 1.5$ assuming $M_t = 175$ GeV.

[a]In this case the decays into charm quarks and gluons also occur at a level of $\sim 5\%$

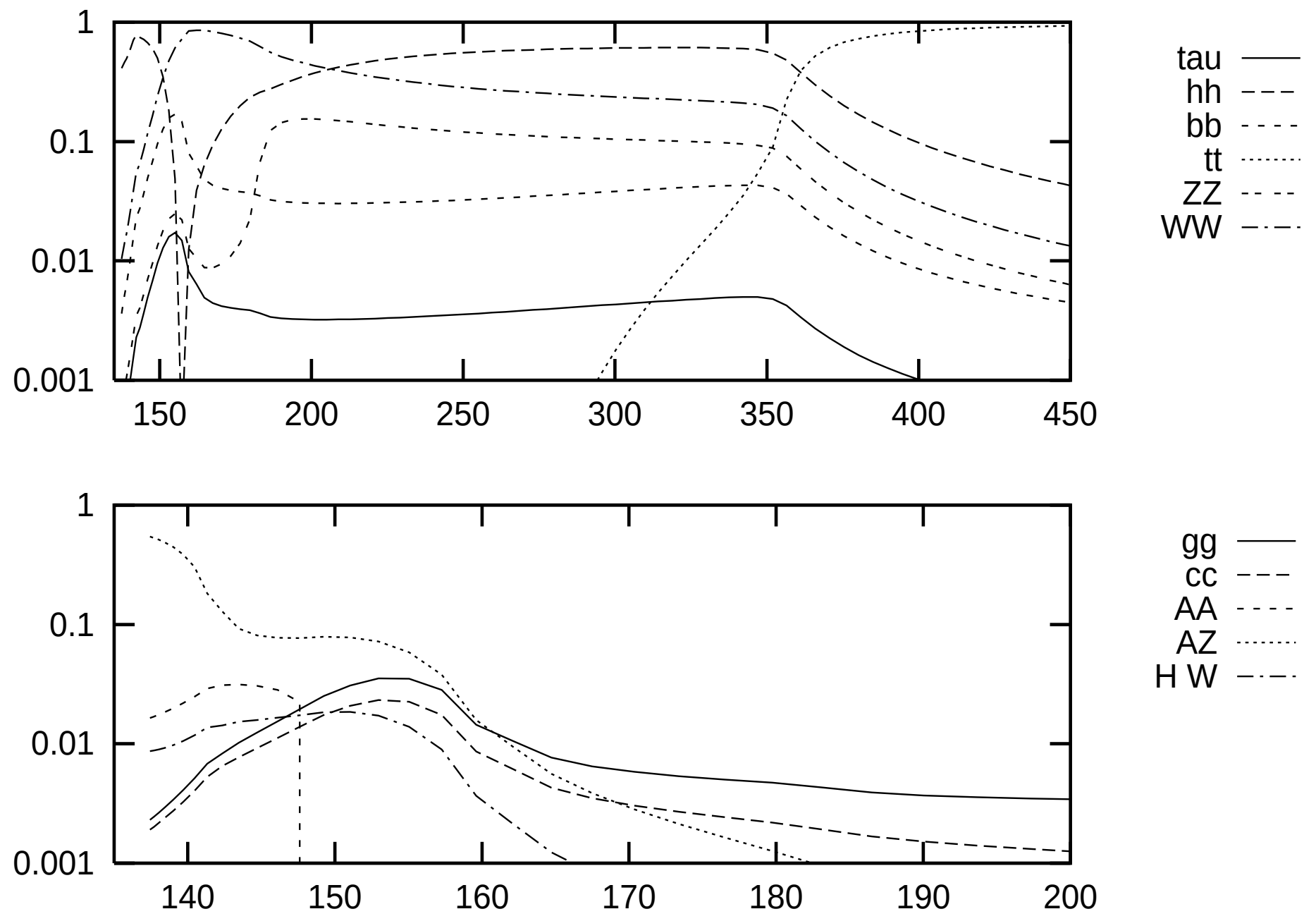

Fig. 1. The branching ratios for the heavier CP-even Higgs boson as functions of $M_H$ for $\tan\beta = 1.5$.

2.1 Three-body decays of the lighter CP-even Higgs boson.

Since its mass is smaller than $\sim 100$ GeV for $\tan\beta \sim 1$, the three-body decays of the lighter Higgs boson $h$ may play some role only in the upper mass range, $M_h \gtrsim 85$ GeV, where $h$ has almost SM-like couplings. In this small mass window the branching rate of the lighter Higgs decay into $WW^*$ can reach the level of a few permille exceeding the physically interesting $h \to \gamma\gamma$ mode which has a branching ratio of $\mathcal{O}(10^{-3})$. Note that the decay $h \to AA$ occurs only for very small values of $M_A$, which are already excluded by LEP data[9].

2.2 Three-body decays of the heavier CP-even Higgs boson

For masses below the $t\bar{t}$ threshold, the $WW/ZZ$ couplings of the heavier CP-even scalar $H$ are not very suppressed for $\tan\beta \sim 1$ and the decays $H \to WW^*/ZZ^*$ below the $2M_{W,Z}$ threshold occur with an appreciable rate, see Fig.1. The decays into two pseudoscalar Higgs bosons $H \to AA$ are restricted to a very small domain in the parameter space and can occur at the level of a few percent, and the on-shell decay $H \to hh$ can reach up 50 % below $t\bar{t}$ threshold. Allowing for one of the final Higgs bosons to be virtual, that is three-body decays $H \to AA^* \to Ab\bar{b}$ or $H \to hh^* \to hb\bar{b}$, dramatically suppresses the rate because for $\tan\beta \sim 1$ the $A$ or $h$ have very small Yukawa couplings to bottom quarks. Another possible three-body decays that we find interesting are the cascade decays $H \to AZ^*$ and $H \to H^{\pm}W^{\mp *}$. Actually, the channel $AZ$ can be open for a large top quark mass which, through radiative corrections, increases $M_H$ to values larger than $M_Z + M_A$. Increasing $M_H$, the mass of $A$ also increases and for $M_H \geq 140$ GeV the decay $AZ$ can proceed only via $Z$ boson off-shell, *i.e.* three-body decay, with the branching ratio $\mathcal{O}(10^{-1})$ until the channel $WW$ is available. The $H \to H^\pm W^{\mp *}$ channel, which is always below threshold, occurs at the level of 1 %. Finally, a few tens of GeV below the $2M_t$ threshold, the decay of $H$ into a real and a virtual top quark starts to be substantial. Above the $2M_t$ threshold, the decay $H \to t\bar{t}$ overwhelms all other decays and there is little chance for the effect of three-body decays to be important.

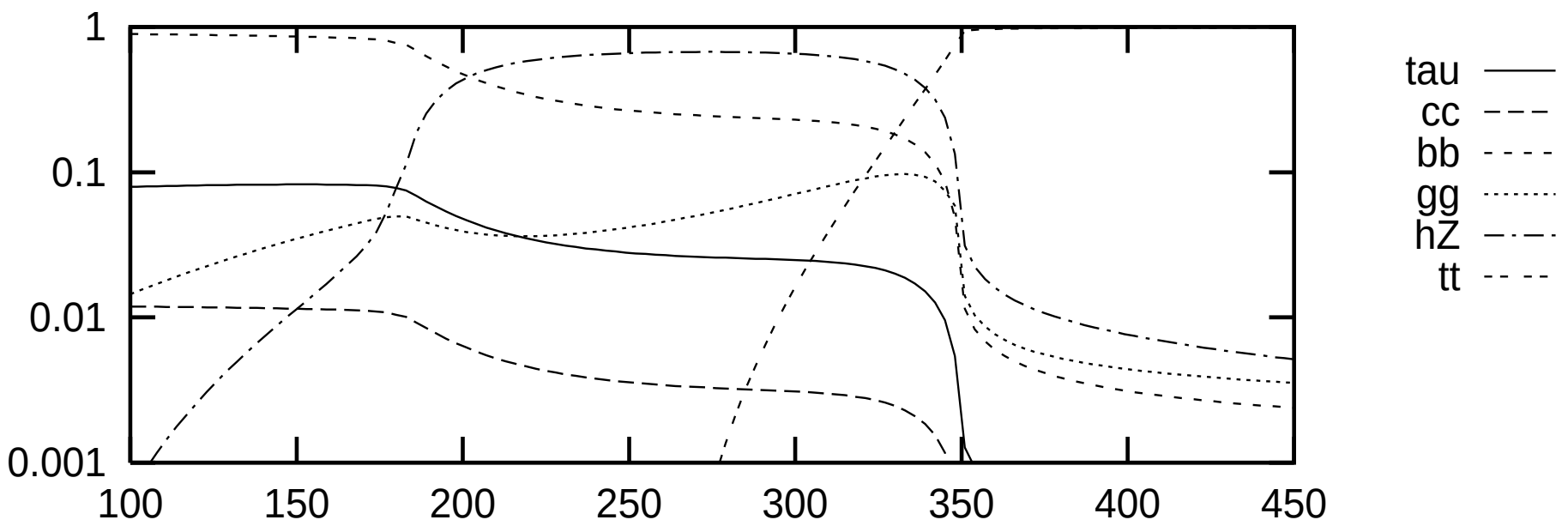

Fig. 2. The branching ratios for the CP-odd Higgs boson as functions of $M_A$ for $\tan\beta = 1.5$.

2.3 Three-body decays of the CP-odd Higgs boson

For the pseudoscalar Higgs boson the situation is much more interesting. Since it has no couplings to the vector bosons, three-body decays involving heavy particles have an important effect, as can be seen in Fig.2. In some mass range, the propagator suppression is compensated (at least partly) by the much larger couplings of $A$ to the heavy states compared to the coupling to the $b$ quarks into which $A$ decays dominantly. This happens in the case of the three-body decay, $A \to hZ^* \to hf\bar{f}$ below 180 GeV, where the price for $Z$ boson to be off-shell and to decay into light fermions is neutralised by the $AZh$ coupling which is much larger than the $Abb$ coupling. On the other hand, the rate for $A \to Zh^*$ is marginal (like in the case $H \to AA^*$ or $H \to hh^*$) because of the extremely small $hbb$ Yukawa coupling for $\tan\beta \sim 1$. The two other possibilities, $A \to HZ^*$ and $A \to H^\pm W^{\mp *}$ are kinematically closed since the pseudoscalar $A$ is always lighter than the $H$ and $H^\pm$ bosons. Finally, the decay $A \to t\bar{t}$ with one of the top quarks being virtual is very important for masses not too far from the threshold region, since the virtuality of the top quark is compensated by the ratio $\sim M_t^2/M_b^2$ of the squared top/bottom Yukawa couplings.

2.4 Three-body decays of the charged Higgs boson

The possibility of three-body decays for the charged Higgs boson $H^+ \to t\bar{b} \to b\bar{b}W^+$ below the top-bottom threshold is even more favorable than in the case of the pseudoscalar $A$, Fig.3. This is due to the smaller value of the $\tau$ lepton and charm quark masses compared to the bottom mass. Similarly, the $H^\pm Wh$ coupling (which is about the same size as the $AZh$ coupling) is much larger than the $H^\pm \tau\nu_\tau$ and $H^\pm cs$ Yukawa couplings, allowing for a sizeable branching ratio for the (off-shell below 160 GeV) decay $H^\pm \to hW^{\pm *}$. Here again the rate for $H^\pm \to Wh^*$ is very small for values of $\tan\beta$ close to unity and the decay $H^\pm \to HW^{\pm *}$ does not occur for real $H$ bosons. The decay $H^\pm \to AW^{\mp *}$ is always off-shell and less phase-space favored than the $H^\pm \to hW^*$. Note that the effect of the three-body decays $AW^*$, $hW^*$ and $\bar{b}t^*$ in the mass range $M_{H^\pm}$ = 90–140 GeV is to lower the decay rate for the most favorable process $\tau\nu$ from $\sim$ 90% to 50–70 %. This may have some impact on the interpretation of the top candidate events in the context of extended models.

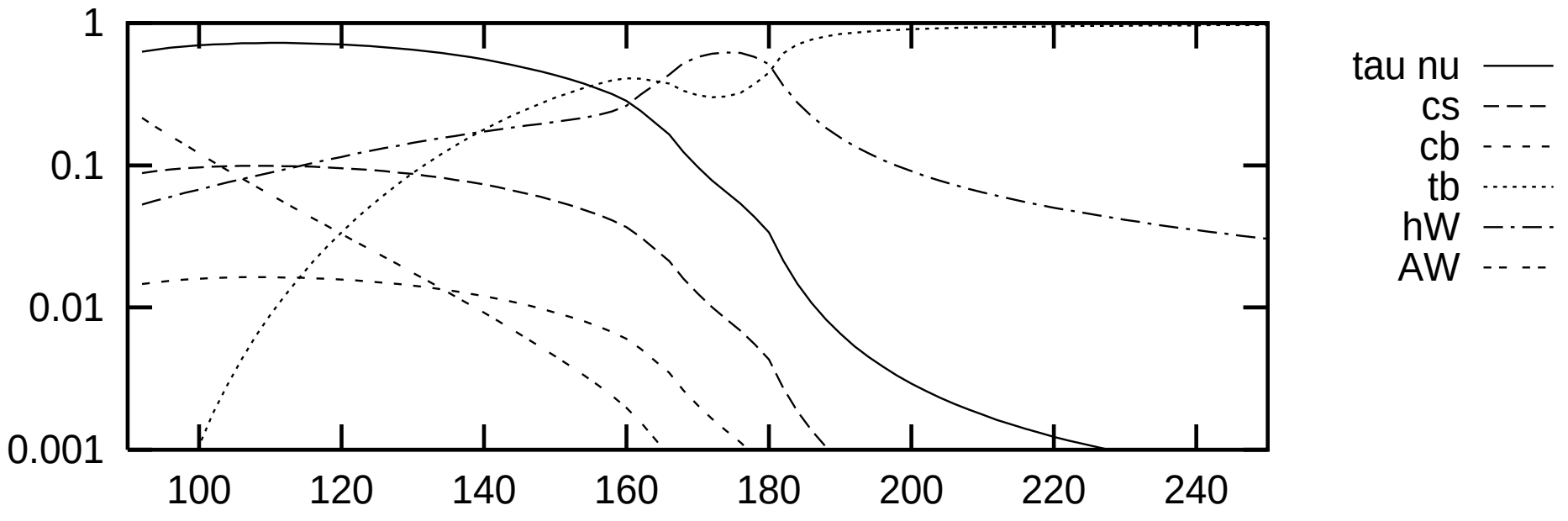

Fig. 3. The branching ratios for the charged Higgs boson as functions of $M_{H^\pm}$ for $\tan\beta = 1.5$.

## 3. Summary

We have found that for small values of $\tan\beta$ the three-body decays of Higgs bosons can be important phenomenologically. In some cases they can reduce significantly the decay rates of dominant two-body processes below their values calculated without taking three-body decays into account.

## 4. Acknowledgements

I would like to express my thanks to A. Djouadi and P. Zerwas for a fruitful collaboration, and A. Ballestrero for several discussions. This work was supported in part by a grant from the Committee for Scientific Research 2 P302 095 05.